%% file: main.tex
\documentclass{article}
\usepackage{spconf,amsmath,graphicx}

\usepackage{booktabs}
\usepackage{balance}

\usepackage{amsmath,epsfig,amssymb,amsthm,url,amsfonts}
\usepackage{algorithm}
\usepackage{algpseudocode}
\usepackage{multirow}
\usepackage{mdframed}
\usepackage{url}
\usepackage{enumitem}
\usepackage[makeroom]{cancel}
\usepackage{dblfloatfix}
\usepackage{array}
\usepackage{comment}
\usepackage{epstopdf}
\usepackage{float}
\usepackage{subfig}
\usepackage{breqn}
\usepackage{cite}
\usepackage{xcolor}
\usepackage{cases}
\usepackage{tabularx}
\usepackage[printonlyused,withpage]{acronym}
\usepackage{balance}
\usepackage{graphbox} 

\usepackage{hyperref}
\setlist[itemize]{label=$\triangleright$}
\newtheoremstyle{break}
{}
{}
{\itshape}
{}
{\bfseries}
{.}
{\newline}
{}
\theoremstyle{break}

\theoremstyle{definition}

\newcommand{\vect}[1]{\mathbf{#1}}
\newcommand{\bs}[1]{\boldsymbol{#1}}
\newcommand{\E}{\mathbb{E}}
\usepackage{url}

\makeatletter
\def\thmhead@plain#1#2#3{%
	\thmname{#1}\thmnumber{\@ifnotempty{#1}{ }\@upn{#2}}%
	\thmnote{ {\the\thm@notefont#3}}}
\let\thmhead\thmhead@plain
\makeatother

\graphicspath{{./figs/}}

\newcommand{\argmax}{\operatornamewithlimits{argmax}}

\input{Header}

\acrodef{SE}{speech enhancement}
\acrodef{STFT}{short-time Fourier transform}
\acrodef{STOI}{short-time objective intelligibility}
\acrodef{PSD}{power spectral density}
\acrodef{NMF}{non-negative matrix factorization}
\acrodef{AV}{audio-visual}
\acrodef{DNN}{deep neural network}
\acrodefplural{DNNs}{deep neural networks}
\acrodef{VAE}{variational auto-encoder}
\acrodefplural{VAEs}{variational auto-encoders}
\acrodef{CVAE}{conditional variational auto-encoder}
\acrodefplural{CVAEs}{conditional variational auto-encoders}
\acrodef{A-VAE}{audio VAE}
\acrodef{V-VAE}{visual VAE}
\acrodef{AV-CVAE}{audio-visual CVAE}
\acrodef{ROI}{region of interest}
\acrodef{MCMC}{Markov Chain Monte Carlo}
\acrodef{EM}{expectation-maximization}
\acrodef{MCEM}{Monte Carlo expectation-maximization}
\acrodef{TF}{time frequency}
\acrodef{ELBO}{evidence lower bound}
\acrodef{ROI}{region of interest}
\acrodef{LR}{Living Room}
\acrodef{SDR}{signal-to-distortion ratio}
\acrodef{PESQ}{perceptual evaluation of speech quality}
\acrodef{ASE}{audio speech enhancement}
\acrodef{VSE}{visual speech enhancement}
\acrodef{AVSE}{audio-visual speech enhancement}
\acrodef{SNR}{signal-to-noise ratio}
\acrodefplural{SNRs}{signal-to-noise ratios}
\acrodef{LSTM}{long short-term memory}
\acrodef{DNNs}{deep neural networks}

\title{Fast and Efficient Speech Enhancement with Variational Autoencoders}
\name{%
Mostafa Sadeghi and
Romain Serizel %
}
\address{%
Université de Lorraine, CNRS, Inria, LORIA, F-54000 Nancy, France}
\begin{document}
%
\maketitle
\begin{abstract}
Unsupervised speech enhancement based on variational autoencoders has shown promising performance compared with the commonly used supervised methods. This approach involves the use of a pre-trained deep speech prior along with a parametric noise model, where the noise parameters are learned from the noisy speech signal with an expectation-maximization (EM)-based method. The E-step involves an intractable latent posterior distribution. Existing algorithms to solve this step are either based on computationally heavy Monte Carlo Markov Chain sampling methods and variational inference, or inefficient optimization-based methods. In this paper, we propose a new approach based on Langevin dynamics that generates multiple sequences of samples and comes with a total variation-based regularization to incorporate temporal correlations of latent vectors. Our experiments demonstrate that the developed framework makes an effective compromise between computational efficiency and enhancement quality, and outperforms existing methods.
\end{abstract}
\section{Introduction}
\label{sec:intro}
Speech enhancement based on deep generative models has attracted much attention recently \cite{pascual2017segan,nugraha2020flow,bando2018statistical,leglaive2018variance,sadeghi2020audio,fang2021variational,carbajal2021guided}. In particular, unsupervised speech enhancement based on variational autoencoder (VAE) \cite{KingW14} consists of a pre-training phase during which a deep speech prior is learned using only clean speech data. At test time, the learned speech prior is used to infer a target clean speech from a given noisy audio recording, e.g., by computing the posterior mean. In this step, a parametric Gaussian noise model is considered whose parameters are learned from the observed noisy speech signal based on \ac{EM} \cite{bando2018statistical,leglaive2018variance}. 

This framework stays in contrast with the supervised methods based on deep learning, which train a deep neural architecture to directly map a given noisy speech signal to a clean version or a time-frequency mask \cite{wang2018supervised}, without explicitly modeling the statistical characteristics of speech signals. On the other hand, unsupervised methods extend the traditional speech enhancement approaches based on linear statistical models, e.g., \ac{NMF}, by incorporating the expressive representation learning frameworks provided by deep neural architectures \cite{lecun2015deep}. As such, they are more interpretable, and could potentially achieve better generalization performance than supervised methods, because noise characteristics are modeled and learned only at test time \cite{bando2018statistical,leglaive2018variance}. 

A disadvantage of unsupervised methods is that they are inefficient at test time, because of the iterative \ac{EM} process. The computational bottleneck comes from the expectation step, which involves an intractable posterior distribution over the latent vectors of the model. This is tackled by a Markov Chain Monte Carlo (MCMC) based sampling method in \cite{leglaive2018variance}. A variational \ac{EM} (VEM) algorithm is proposed in \cite{leglaive2020recurrent} to approximate the posterior with a parametric Gaussian form. Nevertheless, these approaches are computationally expensive. An optimization-based method is also proposed in \cite{kameoka2019supervised,leglaive2020recurrent} that approximates the intractable posterior using only its mode. This, however, leads to performance degradation as the involved posterior is usually multi-modal. An alternative approach is proposed in \cite{pariente2019statistically} which reuses the pre-trained model to approximate the intractable posterior, leading to a fast inference algorithm. However, its performance depends largely on the quality and generalization capability of the learned speech model. 

In this paper, we develop a computationally efficient sampling framework based on Langevin dynamics \cite{welling2011bayesian} to approximate the intractable posterior in the EM phase. This consists of generating multiple sets of samples by using only the gradient information of the log posterior along with injecting stochastic noise to better explore high-density regions of the posterior. Moreover, to incorporate the temporal correlations between consecutive latent vectors, we propose a total variation (TV) based regularization that further improves the performance. The proposed framework makes an effective compromise between the MCMC and optimization-based methods in terms of output quality and computations during the EM phase. Our experiments on speech enhancement confirm that the proposed methodology demonstrates a promising performance and outperforms the previous approaches.

The rest of the paper is organized as follows. Section~\ref{sec:vae} reviews VAE-based speech prior learning and enhancement. The proposed speech enhancement framework is detailed in Section~\ref{sec:prop}. Experimental results are presented in Section~\ref{sec:exp}. Finally, Section~\ref{sec:conc} concludes the paper.

\vspace{-2mm}
\section{Background}
\label{sec:vae}
In this section, we first review the VAE-based speech generative modeling framework and then discuss different approaches to perform speech enhancement using the learned deep speech prior.
\subsection{Speech prior learning}
Speech signals are first transformed to the time-frequency domain by computing the \ac{STFT}, yielding a sequence of time frames denoted $\vect{s}=\{\vect{s}_1,\ldots,\vect{s}_T\}$, where ${\vect{s}_t = [s_{ft}]_{f=1}^F \in \mathbb{C}^F}$. To model the generative process of these time frames, a latent vector, denoted $\vect{z}_t\in \mathbb{R}^L$, is attributed to each time frame $\vect{s}_t$, which encodes the generative information of $\vect{s}_t$ in a lower dimension, i.e.,  ${L\ll F}$. The way $\vect{s}_t$ is generated from $\vect{z}_t$ is modeled by a parametric Gaussian from, i.e., $p_\theta(\vect{s}_t|\vect{z}_t)$, where its mean and variance, as non-linear functions of $\vect{z}_t$, are provided by a so-called decoder network. For the prior of the latent codes, i.e., $p(\vect{z}_t)$, a standard Gaussian distribution is usually assumed. Therefore, the joint distribution of the observed and latent variables can be written as $p_\theta(\vect{s}_t,\vect{z}_t)=p_\theta(\vect{s}_t|\vect{z}_t)p(\vect{z}_t)$.

Having a training set of time frames $\vect{s}$, the next step is to learn the parameters of the generative model, i.e., $\theta$, via EM. This amounts to solving the following problem (M-step):
\begin{equation}\label{eq:training}
\theta^*=\argmax_{\theta}~\sum_{t=1}^T\E_{p_\theta(\vect{z}_t|\vect{s}_t)}\lk\log p_\theta(\vect{s}_t, \vect{z}_t)\rk
\end{equation}
which requires computation of the intractable posterior distributions $p_\theta(\vect{z}_t|\vect{s}_t)$ (E-step). The solution proposed in VAE is to approximate this distribution with a parametric Gaussian form, where, similarly to the decoder, the mean and variance are functions of $\vect{s}_t$ parameterized with a \ac{DNN}, called the encoder, with parameters denoted $\psi$. That is, $q_\psi(\vect{z}_t|\vect{s}_t)\approx p_\theta(\vect{z}_t|\vect{s}_t)$. The encoder and decoder parameters are then jointly learned following a computationally efficient framework that involves optimizing a lower bound on the intractable data log-likelihood $\log p_\theta(\vect{s})$ \cite{KingW14}.
\subsection{Speech enhancement}
Let us denote the \ac{STFT} representation of a given noisy speech signal as $\vect{x}=\{\vect{x}_1,\ldots,\vect{x}_{\Tilde{T}}\}$, where $\vect{x}_t = \vect{s}_t+\vect{b}_t$, with $\vect{b}_t$ corresponding to noise. To complete the observation model, in addition to the pre-trained speech prior, i.e., $p_\theta(\vect{s}_t, \vect{z}_t)$, we need to specify a noise model. A common choice is to consider a circularly symmetric Gaussian form $p_\phi(\vect{b}_t)\sim \mathcal{N}_c(\bs{0}, \diag([\Wb\Hb]_t))$ whose variance is parameterized with an \ac{NMF} model \cite{bando2018statistical}. The two non-negative low-rank matrices, $\Wb,\Hb$, constitute the noise parameters $\phi$. To learn $\phi$ from $\vect{x}$, one would need to solve a similar problem as \eqref{eq:training}, that is
\begin{equation}\label{eq:inference}
\phi^*=\argmax_{\phi}~\sum_{t=1}^{\Tilde{T}}\E_{p_\phi(\vect{z}_t|\vect{x}_t)}\lk\log p_\phi(\vect{x}_t, \vect{z}_t)\rk
\end{equation}
where $p_\phi(\vect{x}_t, \vect{z}_t)=p_\phi(\vect{x}_t| \vect{z}_t)p(\vect{z}_t)$, and the likelihood $p_\phi(\vect{x}_t| \vect{z}_t)$ can easily be computed noting the independence of $\vect{s}_t$ and $\vect{b}_t$ \cite{leglaive2018variance}. Once learned, the speech signal is estimated as the posterior mean $\hat{\vect{s}} = \E_{p_{\phi^*}(\vect{s}|\vect{x})} \lk \vect{s}\rk$ \cite{leglaive2018variance}. Here, too, the posterior $p_\phi(\vect{z}_t|\vect{x}_t)$ in \eqref{eq:inference} is intractable to compute. However, noting the similarity between \eqref{eq:inference} and \eqref{eq:training}, an immediate solution would be to follow the VAE framework and learn an approximate distribution. One could also fine-tune $q_\psi(\vect{z}_t|\vect{s}_t)$ on $\vect{x}_t$, as with the variational \ac{EM} (VEM) approach proposed in \cite{leglaive2020recurrent}. Nevertheless, the VEM method is not computationally efficient, especially when the encoder is very complex.

Another solution is to approximate the intractable expectation in \eqref{eq:inference} by sampling from $p_\phi(\vect{z}_t|\vect{x}_t)\propto p_\phi(\vect{x}_t|\vect{z}_t)p(\vect{z}_t)$ to form a Monte-Carlo (MC) average using the Metropolis-Hastings algorithm \cite{robert1999monte}, resulting in the MCEM method proposed in \cite{leglaive2018variance}. The Metropolis-Hastings algorithm is an iterative MCMC-based method that enables sampling from distributions that are known up to a normalization factor, e.g., $p_\phi(\vect{z}_t|\vect{x}_t)$. However, it could take many iterations to obtain a sequence of samples that resemble those from the true distribution. As such, the resulting MCEM approach might be computationally expensive. A lightweight alternative is to rely on a single sample to approximate the expectation \eqref{eq:inference} by finding the mode of $p_\phi(\vect{z}_t|\vect{x}_t)$, e.g. using a gradient-based solver, leading to the point-estimate EM (PEEM) method  \cite{leglaive2020recurrent}. Nevertheless, given that the posterior is most likely multi-modal, this approach might fail to achieve high-quality results. 

\section{Proposed framework}
\label{sec:prop}
In this section, we present our proposed framework based on Langevin dynamics for approximating the expectation in \eqref{eq:inference} by efficiently sampling from the involved intractable distribution. To this end, we first present a direct application of Langevin dynamics, and then propose a more specialized version by taking into account the specific structure of our problem. The proposed framework combines the advantages of the MCEM and PEEM methods and outperforms them, as will be discussed in Section~\ref{sec:exp}.
\subsection{Langevin dynamics}
With Langevin dynamics, we can generate a sequence of samples from the intractable posterior $p_\phi(\vect{z}_t|\vect{x}_t)$ using only its score function defined as follows:
\begin{equation}
    f(\vect{z}_t)=\nabla_{\vect{z}_t} \log p_\phi(\vect{z}_t|\vect{x}_t) = \nabla_{\vect{z}_t} g(\vect{z}_t)
\end{equation}
where
\begin{equation}
    g(\vect{z}_t) = \log p_\phi(\vect{x}_t|\vect{z}_t) + \log p(\vect{z}_t).
\end{equation}
Given an initial state $\vect{z}_t^{(0)}$, the next states (samples) are produced according to the following rule ($k\ge 0$):
\begin{equation}\label{eq:ld}
    \vect{z}_t^{(k)} = \vect{z}_t^{(k-1)} + \frac{\eta}{2} f(\vect{z}_t^{(k-1)}) + \sqrt{\eta} \bs{\zeta},
\end{equation}
where $\bs{\zeta}\sim\mathcal{N}(0, \Ib)$, and $\eta>0$ is a step size. When $k\rightarrow \infty$ and $\eta \rightarrow 0$, one would have $\vect{z}_t^{(k)}\sim p_\phi(\vect{z}_t|\vect{x}_t)$ under some regularity conditions \cite{welling2011bayesian}. Injecting noise 
$\bs{\zeta}$ into the gradient update prevents the final sample to collapse into the modes and helps better explore high-density regions of the posterior.
\subsection{Extended Langevin dynamics}
In contrast to the original formulation of Langevin dynamics, we propose a more general approach where multiple sequences of samples are obtained in parallel, instead of a single sequence. Starting from $\vect{z}=\lk \vect{z}_{1}, \cdots, \vect{z}_{\Tilde{T}}\rk$, we first draw $m$ different states per each latent vector $\vect{z}_{t}$, denoted $\vect{z}_{t,1},\cdots, \vect{z}_{t,m}$, in a random walk manner by sampling from the following proposal distribution:
\begin{equation}
\label{eq:proposal}
    \vect{z}_{t,i}|\vect{z}_{t} \sim \mathcal{N}(\vect{z}_{t}, \sigma^2\Ib),~~~\forall t, i
\end{equation}
where $\sigma^2>0$ is a given variance parameter. The above sampling can be equivalently written as $\vect{z}_{t,i} = \vect{z}_t + \sigma \bs{\epsilon}$, where $\bs{\epsilon}\sim\mathcal{N}(0, \Ib)$. Compared with \eqref{eq:ld}, this additional level of stochasticity  further helps cover high-density regions.

Furthermore, a direct application of Langevin dynamics would ignore time dependencies (correlations) between the sequences of latent codes $\lk\vect{z}_1,\cdots\vect{z}_{\Tilde{T}}\rk$. To solve this issue, we propose to use a TV-based regularization by extending the original optimization function as follows:
\begin{equation}
    h_\lambda(\vect{z}) = \sum_{t=1}^{\Tilde{T}} g(\vect{z}_t)+ \lambda \sum_{t=2}^{\Tilde{T}} \| \vect{z}_{t}-\vect{z}_{t-1}\|_1
\end{equation}
where $\|.\|$ computes the $\ell_1$ norm of a vector, i.e., the sum of absolute values, and $\lambda\ge 0$ is a trade-off parameter balancing the impact of the TV regularization. This effectively imposes a proximity constraint on the samples of consecutive latent vectors to incorporate their correlations. So, we define
\begin{equation}
    f_\lambda(\vect{z}) = \nabla_{\vect{z}} h_\lambda(\vect{z}),
\end{equation}
which replaces $f(.)$ in \eqref{eq:ld}.

The overall Langevin dynamics (LD) and the EM algorithm to solve \eqref{eq:inference} for speech enhancement (LDEM) are summarized in Alg.~\ref{alg:ld} and Alg.~\ref{alg:ldem}, respectively. Updating $\phi$ in line 9 of Alg.~\ref{alg:ldem} is done using multiplicative update rules \cite{leglaive2018variance}. The original LD method \cite{welling2011bayesian} usually requires many iterations to converge, after which, the initial states corresponding to the so-called burn-in period are discarded. The remaining states could then be used to approximate desired intractable expectations as a weighted Monte-Carlo average. Nevertheless, for our problem, there is no need to run the LD method for a large number of iterations, thanks to the fact that here, the LD iterations are performed inside another iterative process, i.e., LDEM. Therefore, it benefits from warm-starting. 
\begin{algorithm}[t!]
\caption{LD}
\label{alg:ld}
\begin{algorithmic}[1]
\State \textbf{Require:} $\Bar{\vect{z}}^{(0)} = \lk \vect{z}_{t,i}^{(0)}\rk_{t,i}$, $K$, $\eta$, $m$.
\For{$k=1,\cdots,K$}
\State $\bs{\zeta}_{t,i}\sim\mathcal{N}(0, \Ib)$,~~~$\forall t,i$
\State $    \Bar{\vect{z}}^{(k)} = \Bar{\vect{z}}^{(k-1)} + \frac{\eta}{2} f_\lambda(\Bar{\vect{z}}^{(k-1)}) + \sqrt{\eta} \bs{\zeta}$,
\EndFor

\State \textbf{Output:} $\Bar{\vect{z}}^{(K)} = \lk \vect{z}_{t,i}^{(K)}\rk_{t,i}$
\end{algorithmic}
\end{algorithm}

\begin{algorithm}[t!]
\caption{LDEM}
\label{alg:ldem}
\begin{algorithmic}[1]
\State \textbf{Require:} $\vect{x}=\lk \vect{x}_1,\cdots,\vect{x}_{\Tilde{T}}\rk$, $f_\lambda$, $p_\phi$, $\sigma$, $J$.
\State \textbf{Initialize:} $\vect{z}$, $\Wb$, $\Hb$.
\For{$j=1,\cdots,J$}
\For{$t=1,\cdots,\Tilde{T}$ and $i=1,\cdots,m$}
\State $\bs{\epsilon}\sim\mathcal{N}(0, \Ib)$
\State $\vect{z}_{t,i} = \vect{z}_t + \sigma \bs{\epsilon}$
\EndFor
\State $\lk \vect{z}_{t,i}\rk_{t,i} \leftarrow \text{LD}(\lk \vect{z}_{t,i}\rk_{t,i})$
\State $\phi\leftarrow\argmax_{\phi}~\sum_{t,i}\log p_\phi(\vect{x}_t| \vect{z}_{t,i})$
\EndFor

\State \textbf{Output:} $\phi = \lk \Wb, \Hb\rk$
\end{algorithmic}
\end{algorithm}

\begin{table*}[t!]
\centering
	\caption{Average values of the SI-SDR, PESQ, and STOI metrics for the input (unprocessed) and enhanced speech signals.}\vspace{-2mm}
\resizebox{\textwidth}{!}{
\begin{tabular}{|l|c|c|c|c|c||c|c|c|c|c||c|c|c|c|c|}
\hline
 Metric & \multicolumn{5}{c||}{SI-SDR (dB)} & \multicolumn{5}{c||}{PESQ} & \multicolumn{5}{c|}{STOI} \\
\hline
{Noise SNR (dB)} & {-10} & {-5} & {0} & {5} & {10} & {-10} & {-5} & {0} & {5} & {10} & {-10} & {-5} & {0} & {5} & {10} \\ \hline\hline
Input (unprocessed) & -18.08 & -12.80 & -7.72 & -2.91  & 2.04    & 1.40 & 1.51 & 1.76 & 2.05 & 2.37 & 0.12        & 0.20       & 0.30       & 0.43       & 0.56                    \\ \hline\hline
PEEM \cite{leglaive2020recurrent} & -9.66 & -4.35 & 0.57 & 5.49 & 10.33 & {1.60} & 1.80 & 2.06 & 2.36 & 2.67 & 0.15         & 0.24        & 0.36      & 0.49       & 0.63 \\ \hline
MCEM \cite{leglaive2018variance}  & {-7.67} & {-1.48} & {3.34} & {7.81} & {12.00} &  {1.55} & {1.84} & {2.18} & {2.49} & {2.78} & {\textbf{0.17}} & {\textbf{0.27}} & {\textbf{0.40}} & {\textbf{0.54}} & {\textbf{0.66}}\\ \hline\hline
LDEM {\footnotesize ($\lambda:0,m:1$)} & \textbf{-7.20} & \textbf{-1.03} & \textbf{3.76} & \textbf{8.18} & \textbf{12.37}  & {1.54} & {1.85} & {2.18} & {2.50} & {2.78} & 0.16                    & 0.25                  & 0.38                     & 0.52                     & 0.65 \\ \hline
LDEM {\footnotesize ($\lambda:0.5,m:1$)}& {-7.17} & {-1.08} & {3.70} & {8.16} & {12.34}  & {1.58} & {1.87} & {2.20} & {2.51} & {2.80} & 0.17                    & 0.27                  & 0.40                     & 0.53                     & 0.66 \\ \hline
LDEM {\footnotesize ($\lambda:5,m:1$)}& {-7.28} & {-1.41} & {3.42} & {7.93} & {12.13}  & {1.70} & {1.96} & {2.25} & {2.56} & {2.83} & {0.17}                    & {0.27}                  & {0.40}                     & {0.53}                     & {0.66} \\ \hline

LDEM {\footnotesize ($\lambda:5, m:5$)}&  {-7.10} & {-1.26} & {3.60} & {8.07} & {12.27} & \textbf{1.73} & \textbf{2.01} & \textbf{2.30} & \textbf{2.59} & \textbf{2.85} & \textbf{0.17}                    & \textbf{0.27}                  & \textbf{0.40}                     & \textbf{0.54}                     & \textbf{0.67} \\ \hline

\end{tabular}}
\label{tab:se_results}\vspace{-3mm}
\end{table*}
\begin{table}[t!]
\centering
	\caption{Average runtimes (in seconds) of different speech enhancement methods per test sample ($\sim$ 5-second long).}\vspace{-2mm}
\resizebox{0.48\textwidth}{!}{
\begin{tabular}{|l|c|c|c|c|}
\hline
Method & PEEM \cite{leglaive2020recurrent} & MCEM \cite{leglaive2018variance} & LDEM {\footnotesize ($m:1$)} & LDEM {\footnotesize ($m:5$)}                    \\ \hline
runtime  & {5} & 32 & 5.4 & 18 \\
\hline
\end{tabular}}
\label{tab:runtime}\vspace{-3mm}
\end{table}

\section{Experiments}
\label{sec:exp}
\noindent\textbf{Baselines}. In this section, we evaluate the performance of the proposed LDEM framework for speech enhancement, and compare it with the PEEM and MCEM methods \cite{leglaive2020recurrent}. We used the publicly available PyTorch implementations of the two latter methods\footnote{\url{https://gitlab-research.centralesupelec.fr/sleglaive/icassp-2020-se-rvae}}, and implemented LDEM based on that. We also tested the VEM method of \cite{leglaive2020recurrent} using its available implementation, but, unfortunately, it did not work. The poor performance of VEM was already observed in \cite{leglaive2020recurrent}. Furthermore, there is no public code for the method of \cite{pariente2019statistically}, and our own implementation did not work. Therefore, we excluded these two methods from the baselines. 

\vspace{0.3cm}
\noindent\textbf{Evaluation metrics}. The speech enhancement performance is measured based on the standard metrics, including the short-term objective intelligibility (STOI) measure~\cite{Taal2011stoi}, ranging in $[0,1]$, the perceptual evaluation of speech quality (PESQ) score~\cite{Rix2001pesq}, ranging in $[-0.5,4.5]$, and the scale-invariant signal-to-distortion ratio (SI-SDR) \cite{le2019sdr} in dB. Moreover, as a rough measure of the computational complexity of different methods, we report the average runtime in seconds. Our experiments were performed on an Intel Xeon Gold 6130 CPU.

\vspace{0.3cm}
\noindent\textbf{VAE architecture}. The architecture of the VAE considered in our experiments follows the one proposed in \cite{leglaive2018variance}, which consists of an encoder and decoder each having a 128-node single fully-connected hidden layer with hyperbolic tangent activation functions. The dimension of the latent space is set to $L=32$. 

\vspace{0.3cm}
\noindent\textbf{Datasets}. To train the VAE model, we used the speech data in the TCD-TIMIT corpus \cite{harte2015tcd}, which contains speech utterances from 56 English speakers (39 for training, 8 for validation, and 9 for testing) with an Irish accent. There are 98 different audio files per speaker, each with an approximate length of 5 seconds, and sampled at 16 kHz. This results in about 8 hours of data. The STFT of the speech data is computed with a 1024 samples-long (64 ms) sine window, 75$\%$ overlap, and without zero-padding, yielding STFT frames of length $F=513$.

To evaluate the speech enhancement performance, we used some pre-computed noisy versions of the TCD-TIMIT data presented in \cite{abdelaziz2017ntcd}. This includes six different noise types, namely \textit{\ac{LR}}, \textit{White}, \textit{Cafe}, \textit{Car}, \textit{Babble}, and \textit{Street}, with five noise levels: $-10$~dB, $-5$~dB, $0$~dB, $5$~dB, and $10$~dB. For each noise type and noise level, we randomly selected 5 audio files from each test speaker, which resulted in a total of 1350 test files.

\vspace{0.3cm}
\noindent\textbf{Parameters setting}. The VAE model was trained with a batch size of 128 using the Adam optimizer~\cite{KingmaB15}, with a learning rate of $0.0001$. We used early stopping on the validation set with patience of 20 epochs (i.e., training stops if the validation loss does not improve after 20 epochs). The number of EM iterations is set to $J=100$ for all the methods. The MCEM algorithm was run using the default setting considered in \cite{leglaive2020recurrent}. The number of iterations in the E-step has been set to $K=10$ for both PEEM and LDEM. The Adam optimizer has been used for PEEM, as in \cite{leglaive2020recurrent}, with a learning rate of $0.005$. For LDEM, we set $\eta=0.005$ and $\sigma^2=0.01$. We have tried different values for the regularization parameter, $\lambda$, and the number of samples, $m$.

\vspace{0.3cm}
\noindent\textbf{Results}. The speech enhancement results per noise level are reported in Table~\ref{tab:se_results}. The average runtimes of different methods per each test sample ($\sim$ 5-second long) are reported in Table~\ref{tab:runtime}. We can draw several conclusions by inspecting the results. First, it can clearly be seen that LDEM, even without TV regularization and with $m=1$, significantly outperforms PEEM in all the metrics, e.g., {2.7 dB} average performance gain in SI-SDR, with approximately the same runtime. Moreover, LDEM achieves approximately the same PESQ scores as MCEM, but outperforms it in terms of SI-SDR, while having a much lower runtime. However, the STOI scores of LDEM are lower than those of MCEM. 

As we increase $\lambda$ (TV regularization parameter), we see a clear improvement, such that for $\lambda=5$, LDEM outperforms MCEM in terms of PESQ, while showing similar performance in terms of SI-SDR and STOI. Again, LDEM achieves this with a much lower runtime. Furthermore, increasing the number of samples, $m$, causes a performance boost, such that, with a lower runtime, LDEM demonstrates better enhancement metrics than MCEM.

\section{Conclusions}\label{sec:conc}
In this paper, we addressed the EM step of speech enhancement based on VAEs, which involves approximating an intractable latent posterior distribution. While existing approximating methods suffer either from high computational complexity or low-quality output results, we developed an efficient sampling-based method that effectively compromises the complexity and quality. More precisely, the proposed framework builds on Langevin dynamics and extends it to have multiple sets of samples as well as a total variation regularization to take into account the temporal correlations of latent vectors. Our experimental results showed that the proposed method outperforms the previous sampling-based approaches.
\section{Acknowledgement}
Experiments presented in this paper were carried out using the Grid'5000 testbed, supported by a scientific interest group hosted by Inria and including CNRS, RENATER, and several Universities as well as other organizations (see https://www.grid5000.fr). 

\bibliographystyle{IEEEbib}
\bibliography{mybib}

\end{document}

%% file: Header.tex
\newcommand{\lk}{ \left\{ }
\newcommand{\rk}{ \right\} }

\newcommand{\Hb}{{\bf H}}

\newcommand{\Wb}{{\bf W}}

\newcommand{\diag}{\mbox{{diag}}}

\newcommand{\Ib}{{\bf I}}

\newsavebox\mybox

